\documentclass[useAMS, usenatbib]{mnras}

\usepackage{epsfig}
\usepackage{graphics}
\usepackage{amssymb, amsmath, floats, placeins}

\title[Redshift-space multipoles for a masked density field]
      {Rapid modelling of the redshift-space power spectrum multipoles for a masked density field}
\author[Wilson et al.]{M.J. Wilson$^1$ \thanks{email: mjw@roe.ac.uk}, J.A. Peacock$^1$, A. N. Taylor$^1$ and S. de la Torre$^2$ \\
 $^1$Institute for Astronomy, University of Edinburgh, Royal Observatory,
     Blackford Hill, Edinburgh EH9 3HJ\\
 $^2$Aix Marseille Universit\'e, CNRS, LAM (Laboratoire d'Astrophysique de Marseille), UMR 7326, 13388, Marseille, France \\
}

\date{Accepted xxx Received xxx in original form xxx}
\date{\today}
\pagerange{\ref{firstpage}--\ref{lastpage}} \pubyear{2016}

\newcommand{\mpcoh}{\,\ensuremath{h^{-1}}\textrm{Mpc}}
\newcommand{\hompc}{\,\ensuremath{h}\textrm{Mpc}^{-1}}

\def\[{\begin{equation}}
\def\]{\end{equation}}
\def\citejap#1{\citeauthor{#1} \ \citeyear{#1}}
\def\gsim{\mathrel{\lower0.6ex\hbox{$\buildrel {\textstyle >}\over {\scriptstyle \sim}$}}}
\def\lsim{\mathrel{\lower0.6ex\hbox{$\buildrel {\textstyle <}\over {\scriptstyle \sim}$}}}
\def\deg{^\circ}

\begin{document}
\maketitle
\label{firstpage}
\begin{abstract}
In this work we reformulate the forward modelling of the redshift-space power spectrum multipole moments for a masked density field, as encountered in galaxy redshift surveys.  Exploiting the symmetries of the redshift-space correlation function, we provide a masked-field generalisation of the Hankel transform relation between the multipole moments in real and Fourier space.  Using this result, we detail how a likelihood analysis requiring computation for a broad range of desired $P(k)$ models may be executed $10^3-10^4$ times faster than with other common approaches, together with significant gains in spectral resolution.  We present a concrete application to the complex angular geometry of the VIPERS PDR-1 release and discuss the validity of this technique for finite-angle surveys. 
\end{abstract}


\section{Introduction}
\label{sec:intro}
For a Fourier-based analysis of a galaxy redshift survey, the imprint of the survey geometry is commonly the largest systematic difference between the large-scale observed power spectrum and that predicted by fundamental physics.  This difference is of increased importance in a redshift-space distortion (RSD) analysis as the principal observable is the power spectrum anisotropy, which is equally sensitive to the density field and survey mask.  The density field appears anisotropic when inferred from observed redshifts as the radial component of the peculiar velocity field superimposes an additional Doppler shift.  These peculiar velocities are a consequence of the formation of large-scale structure via gravitational collapse, and as such the magnitude of this effect is dependent on the effective strength of gravity on cosmological scales \citep{Gigi2008}.  A measurement of the observed anisotropy by surveys such as Euclid \citep{Euclid} will therefore provide a stringent test of modified gravity theories in the future.

The goal of such surveys is to achieve a statistical error of $\simeq 1 \%$ on the logarithmic growth rate of density fluctuations, which requires systematic uncertainties to be very well understood.  In this work we propose a new approach for the forward modelling of the systematic change in the galaxy power spectrum due to the survey mask.  Because statistical noise is amplified by deconvolution, a forward modelling of this effect as part of a likelihood analysis is the logical approach; but with a broad range of theoretical models to consider it is important to do so efficiently.  We show that the symmetries of the redshift-space correlation function make it possible to do so in a manner that offers a greater physical insight, together with significant gains in speed and resolution when compared to other common methods.
 
In the usual approach the density field as observed by a redshift survey is ``the infinite sea of density fluctuations'' multiplied by a mask $W(\mathbf{x})$ that accounts both for the survey geometry and a local weighting, which may include incompleteness corrections or FKP weights \citep{FKP}:
\[
\delta'(\mathbf{x}) = \delta(\mathbf{x}) \ W(\mathbf{x}).
\label{delta_prime}
\]
This multiplication in configuration space results in a convolution in Fourier Space: $\tilde \delta' (\mathbf{k}) = \tilde \delta (\mathbf{k}) * \ \tilde W(\mathbf{k})$, which, in the absence of phase correlations between density field and the mask, is also true of the observed power \citep{PeacockNicholson}:
\[
\label{P_cnvld}
P'(\mathbf{k}) = \int  \frac{d^3 q}{(2 \pi)^3}  \ P(\mathbf{k} - \mathbf{q}) \ | \tilde W(\mathbf{q})|^2.
\]
Due to statistical homogeneity, different Fourier modes are uncorrelated: $\langle \tilde \delta(\mathbf{k}) \tilde \delta^*(\mathbf{k'}) \rangle = (2 \pi)^3 \delta^3(\mathbf{k} - \mathbf{k}') \, P(\mathbf{k})$, but this is no longer true of $\langle \tilde \delta'(\mathbf{k}) \tilde \delta' {}^*(\mathbf{k'}) \rangle $ \citep{Hamilton06}.  However, the diagonal term is the quantity that contains the cosmological information and it is the systematic change in shape of this function that we seek to calculate.  We adopt a convention in which $P(k)$ has units of volume and exploit the independence of $P'(\mathbf{k})$ on the phases of the density field, by presenting various test cases based on Gaussian random fields.  

This convolution alters both the amplitude and shape of the observed power spectrum with respect to that of the true field.  We assume the amplitude of the observed power has been suitably corrected, by dividing by a factor of
$\int d^3 x  \ W^2(\mathbf{x})
\label{ampcorr}$, and address solely the change of shape in the forward modelling.

As first described by \cite{Kaiser}, $P(\mathbf{k})$ is anisotropic about the line-of-sight when the radial comoving position of a galaxy is inferred from a measured redshift.  On large scales this anisotropy is dependent on the infall (outflow) rate of galaxies into (out of) over (under) densities and hence is sensitive to the strength and therefore theory of gravity on cosmological scales \citep{Gigi2008}.  With an additional large-$k$ suppression due to the virialised motions of galaxies in groups and clusters, a commonly assumed model for the power spectrum is the dispersion model, which combines the Kaiser anisotropy factor with a `fingers-of-God' damping:
\[
P(\mathbf{k}) = \frac{(1 + \beta \mu^2)^2}{1 + \frac{1}{2} k^2 \sigma_p^2 \mu^2} P_{\text{R}}(k).
\label{DispersionModel}
\]
Here $P_R(k)$ is the configuration-space spectrum, $\mu = \mathbf{\hat{k}} \cdot \hat{\boldsymbol \eta}$ for a unit vector $\hat{\boldsymbol \eta}$ lying along the line-of-sight, $\sigma_p$ is an empirical pairwise dispersion (in this case for a Lorentzian damping model) and $\beta$ is the ratio of the logarithmic growth rate of density fluctuations to linear galaxy bias.  For an outline of the approximations underlying this model see e.g \cite{Cole94} and \cite{Cole}.  Taking $\hat{\boldsymbol \eta}$ as the polar axis, the azimuthal symmetry and $\mu^2$ dependence of RSD allows $P(\mathbf{k})$ to be distilled into a series of multipoles, of even order in $\ell$:
\[
\label{legenSeries}
P(\mathbf{k}) = \sum_{\ell =0, 2, 4, \cdots}^{\infty} P_{\ell}(k) L_{\ell}(\mu).
\]
Here $L_{\ell}$ is a Legendre polynomial of order $\ell$; these form a complete basis for $-1 \leq \mu \leq 1$.  The monopole and quadrupole modes are given by $L_0 = 1$ and $L_2 = \frac{1}{2} (3\mu^2 -1)$.  

Although we have presented the symmetries we exploit in the context of the dispersion model, they are inherent to all models that assume the validity of the `distant observer' approximation -- when the variation of the line-of-sight across the survey is neglected and subsequently both $\hat{\boldsymbol \eta}$ and $\mu$ are well defined.  However, this is only valid for surveys of relatively small solid angle or for RSD analyses that are restricted to pairs of small angular separation.  We comment further on the applicability of our approach to finite-angle surveys in \S\ref{wide-angle}.

Despite being a physically well-motivated approximation this dispersion model fails to incorporate the more involved aspects of RSD.  More developed models that apply appropriate corrections include those by \cite{Scoccimarro} and \cite{Taruya}; see \cite{sylvainModels} for further details.  The former relaxes the assumption of linear theory relations between the overdensity and velocity divergence fields on the largest scales surveyed, yielding an effective Kaiser factor that is typically calibrated with numerical simulations, e.g. \cite{Jennings}.  However, the ansatz proposed by \cite{Scoccimarro} continues to neglect the physical origin of the fingers-of-God damping -- the virialised motion of galaxies is sourced by the same velocity field responsible for the linear component of RSD.  \cite{Taruya} apply further corrections that more accurately account for the correlation of this non-linear suppression with the linear velocity divergence field.  Despite these shortcomings, the dispersion model is sufficient for illustration as these more developed models satisfy the symmetries that are exploited.

To implement our approach it is necessary to be able to compute the multipole moments for the assumed RSD model, which are quoted for the Kaiser-Lorentzian model in Appendix \ref{sec:appA}.  In the following section we present the main result of this paper: a masked-field generalisation of the known Hankel transform relation between the multipole moments in real and Fourier space, allowing for the rapid prediction of $P'_{\ell}(k)$ by 1D FFT.

\section{Power spectrum multipoles for a masked density field}
\subsection{Outline of the method}
\label{sec:multipoles}
We start with the convolution:
\[
\label{convo}
P' (\mathbf{k}) =   \ \int  \frac{ d^3 q}{(2 \pi)^3} \ P(\mathbf{q}) \ |\tilde W(\mathbf{k} - \mathbf{q})|^2 . 
\]
The simplest methods to evaluate this integral are by approximating it as a Riemann sum or by the application of the convolution theorem, together with 3D FFTs.  These approaches share a number of disadvantages for inclusion in a likelihood analysis, which is especially the case for the common case of a pencil-beam geometry: 
\begin{enumerate}
\item The broad extent of the $|\tilde W(\mathbf{k})|^{2}$ kernel when the mask is narrow along one or more dimensions in configuration space, such as for the pencil-beam geometry common to $z \simeq 1$ surveys.  In this case a Riemann sum is prohibitively slow as the addition of a large number of non-negligible terms is required.  This calculation must then be repeated for each mode for which $P'(\mathbf{k})$ is desired.
\item Systematic errors introduced by the use of a FFT for the estimation of $|\tilde W(\mathbf{k})|^{2}$, not least due to memory limited resolution and the subsequent aliasing effect (see \S\ref{sec:W2delta} for further discussion).  While these artefacts are also a concern when measuring the multipole moments of the data, this is less of a concern as this measurement needs to be performed once typically.  In this case, Jenkins's folding \citep{Jenkins} may be conveniently applied, which achieves a finer resolution by calculating multiple FFTs, once for each fold.  See \cite{Halofit} and Wilson et al. (in prep.) for additional discussion. 
\item The estimation of the multipole moments of $P'(\mathbf{k})$:  the modes available from a FFT lie on a Cartesian lattice and as such are irregularly spaced in $\mu$.  Therefore it is invalid to calculate the multipole moments by approximating the multipole decomposition, eqn. (\ref{multipoleDecomp}), as a Riemann sum and instead linear regression may be used; see Wilson et al. (in prep.) for a more detailed discussion.  We find that the time required for this decomposition is at least comparable to that spent on a 3D FFT estimate of $P'(\mathbf{k})$.  As it is necessary to perform this calculation for a range spanning many decades in wavenumber, a large number of modes and hence a computationally expensive 3D FFT is required.
\item The necessary calculation of $P'_{\ell}(k)$ for each of a broad range of models in a likelihood analysis, which has led to recent approaches utilising pre-computed lookup tables to optimise the calculation, e.g. \cite{Blake}.  While this is an equally rapid approach, this `mixing matrix' technique has limited portability -- for a given matrix, the model power spectra must be provided in pre-defined wavenumber bins and there is a hard ceiling to the highest order $P_{\ell}'(k)$ that may be calculated before a new matrix is required.  In contrast to this, we show that the multipole moments of the mask autocorrelation function allow for the calculation of $P_{\ell}'(k)$ at any order, with no restrictions on the wavenumber range and resolution. 
\end{enumerate}

In this work we present a reformulation that predicts $P'_{\ell}(k)$ directly; this allows for a rapid implementation that requires only a small number of 1D FFTs per model and hence achieves a significantly greater spectral resolution, which minimises the number of FFT based artefacts. To begin with, we first generalise the known Hankel transform relation between the multipole moments in configuration and Fourier space,
\begin{eqnarray}
\label{Hankelpair}
P_{\ell}(k) &=& 4 \pi (-i)^\ell \int \Delta^2 d\Delta \ \xi_{\ell}(\Delta) \, j_{\ell}(k\Delta),
\end{eqnarray}
to a masked field, by exploiting the symmetries of the redshift-space correlation function. 

Due to the convolution theorem, the autocorrelation functions of both density field, $\xi(\mathbf{\Delta})$, and mask, $Q(\mathbf{\Delta})$, multiply to give the masked autocorrelation:
\[
\label{overautocorr}
\xi'(\mathbf{\Delta}) = \xi (\mathbf{\Delta}) \ Q(\mathbf{\Delta}).
\]
Note that this masked autocorrelation is equally sensitive to the anisotropy of the redshift-space density field and the survey mask.  Here we have introduced both $\xi(\mathbf{\Delta})$ and $Q(\mathbf{\Delta})$ as the inverse Fourier transforms of $P(\mathbf{k}$) and $|\tilde W(\mathbf{k})|^{2}$ respectively,
\[
\label{FTwindow}
Q(\boldsymbol \Delta)  = \int d^3 x \ W(\mathbf{x}) \ W(\mathbf{x} + \boldsymbol \Delta) = \int \frac{d^3 k}{(2\pi)^3} \ |\tilde W(\mathbf{k})|^{2} \ e^{i \mathbf{k} \cdot \boldsymbol \Delta}. 
\]
Spherical coordinates present a natural coordinate system for RSD in which the physical symmetries are best represented. A coordinate transformation may be achieved by expanding each plane wave in spherical waves using the Rayleigh plane wave expansion:
\[
\label{Rayleigh}
e^{-i  \mathbf{k} \cdot \mathbf{\Delta}} = \sum_{p=0}^{\infty} (-i)^{p} (2p + 1) \ j_p(k \, \Delta) L_p(\hat{\mathbf{\Delta}} \cdot \mathbf{\hat{k}}),
\]
which is reproduced from equation (B3) of \cite{Cole94}.  Here $j_{p}(k \, \Delta)$ denotes a spherical Bessel function of order $p$.  Conventionally the chosen observables are the multipole moments of $P(\mathbf{k})$, despite the convolution:
\[
\label{multipoleDecomp}
P'_{\ell}(k) = \frac{(2\ell + 1)}{2} \int d(\hat{\mathbf{k}} \cdot \hat{\boldsymbol \eta}) \int \frac{d \phi_k}{(2 \pi)} P'(\mathbf{k}) \ L_{\ell} (\mathbf{\hat{k}} \cdot \hat{\boldsymbol \eta}), 
\]
for an azimuthal coordinate of $\mathbf{k}$ given by $\phi_k$.  With equations (\ref{overautocorr}) -- (\ref{multipoleDecomp}) we find:
\[
P'_{\ell}(k) = (-i)^{\ell} (2 \ell + 1) \int d^{3} \Delta  \ j_{\ell}(k \Delta) \ \xi' (\mathbf{\Delta})  L_{\ell} (\mathbf{\hat{\Delta}} \cdot \hat{\boldsymbol \eta}).
\label{noangleavg}
\]
Here we have used the relation outlined in Appendix B (see also eqn. A11 of \citejap{Cole94}):
\[
\frac{(2\ell + 1) }{2}\int d(\mathbf{\hat{k}} \cdot \hat{\boldsymbol \eta} ) \int \frac{d \phi_k}{(2 \pi)} L_{\ell}(\mathbf{\hat{k}} \cdot \hat{ \boldsymbol \eta}) \ L_{\ell'} (\mathbf{\hat{k}} \cdot \mathbf{\hat{\Delta}} ) = \delta^{K}_{\ell \ell'} \ L_{\ell} (\mathbf{\hat{\Delta}} \cdot \hat{\boldsymbol \eta} ).
\label{eqn:Orthogonality}
\]
This identity ensures only the $p=\ell$ term survives from the Rayleigh plane wave expansion.  

In the distant observer approximation, when the expansion of $P(\mathbf{k})$ according to eqn. (\ref{legenSeries}) is valid, $\xi(\boldsymbol \Delta)$ may be similarly decomposed and the expression may be further simplified:
\begin{align}
P'_{\ell}(k) = 4 \pi &(-i)^{\ell} \left ( \frac{2 \ell + 1}{2q + 1} \right ) \nonumber \\ 
		& \times A_{\ell, \ell'}^{q} \int \Delta^2 d\Delta \ \xi_{\ell'} (\Delta) \ Q_q (\Delta) \ j_{\ell}(k \Delta).
\label{cnvldpk}
\end{align}
The structure of eqn. (\ref{cnvldpk}) is clear as the multipole moments in configuration and Fourier space form a Hankel transform pair; eqn. (\ref{cnvldpk}) generalises eqn. (\ref{Hankelpair}) to the case of a masked density field and is a primary result of this paper.  For the masked case, the Hankel transform relation is preserved and $\xi_\ell$ is simply replaced by an effective $\xi'_{\ell}$ defined below.  We assume the Einstein summation convention over the repeated \textit{dummy} indices in eqn (\ref{cnvldpk}).  The multipole moments of the mask autocorrelation function have been defined as 
\[
Q_{q}(\Delta) = \left ( \frac{2q + 1}{2} \right ) \int d(\mathbf{\hat{\Delta}} \cdot \hat{\boldsymbol \eta}) \ \int \frac{d\phi_{\Delta}}{(2 \pi)} \ Q(\mathbf{\Delta}) \ L_{q}(\mathbf{\hat{\Delta}} \cdot \hat{\boldsymbol \eta}). 
\]
Finally, Legendre polynomials have been used as a basis in $\mathbf{\hat{\Delta}} \cdot \hat{\boldsymbol \eta}$: 
\[
L_{\ell}(\mathbf{\hat{\Delta}} \cdot \hat{\boldsymbol \eta}) L_{\ell'}(\mathbf{\hat{\Delta}} \cdot \hat{\boldsymbol \eta}) = \sum^{\min(\ell, \ell')}_{q=0} A_{\ell, \ell'}^{q} \ L_{q}(\mathbf{\hat{\Delta}} \cdot \hat{\boldsymbol \eta}).  
\]
The product of two Legendre polynomials has been derived in \cite{PSP:1736752}; this result,
\[
L_{\ell} L_{\ell'} = \sum_{p=0}^{\min(\ell, \ell')} \frac{G_{\ell-p} G_p  G_{\ell'-p}}{G_{\ell+\ell'-p}} \left ( \frac{2\ell + 2\ell' -4p +1}{2\ell + 2\ell' -2p +1} \right) L_{\ell+\ell' -2p},
\label{Bailey}
\]
where 
\[
G_p = \frac{1 . 3 . 5 \dots (2p-1)}{p!} \equiv \frac{2^p (\frac{1}{2})_p}{p!} \text{ and } \ell \geq \ell',
\]
may be used to obtain the $A_{\ell, \ell'}^q$ coefficients.
\begin{figure*}
\centering
\includegraphics[width=0.9\textwidth]{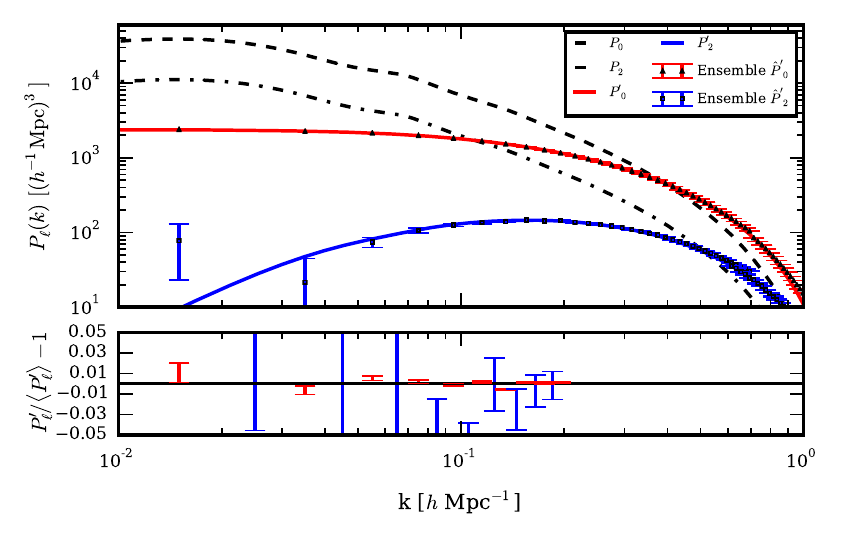}
\caption{A comparison of the power spectrum multipole moments for a masked density field, as predicted by this approach, and the mean of 5000 realisations of an illustrative example comprised of a mask and density field given by 3D Gaussian random fields.  For the mask and overdensity field realisations a $(1 + \mu^2/2)P_R(k)$ model was assumed with $P_{\text{R}}(k)$ taken as an example non-linear $\Lambda$CDM power spectrum for biased tracers, smoothed by a spherical filter of $3 \mpcoh$ in radius.  The step in the standard error of the mean at $k=0.7 \hompc$ corresponds to replacing the results with those obtained from a box of half the length, allowing a higher mesh resolution to be obtained and thus avoiding the spurious increase of power due to aliasing from the finite FFT grid.  The convergence of the convolved to unconvolved multipoles at large $k$ occurs for $k>1 \hompc$, as a result of the strict mask applied, which differs from a realistic survey in that $W(\mathbf{x})$ differs from unity over the majority of the volume.  In the bottom panel is shown a plot of the residuals, which shows consistency is achieved within the measured variance up to $k=0.2 \hompc$; above these scales aliasing from the finite 3D FFT grid introduces a systematic bias at this gridding resolution.}
\label{ConvldMultipoles}
\end{figure*}
\begin{figure*}
\centering
\includegraphics[width=\textwidth]{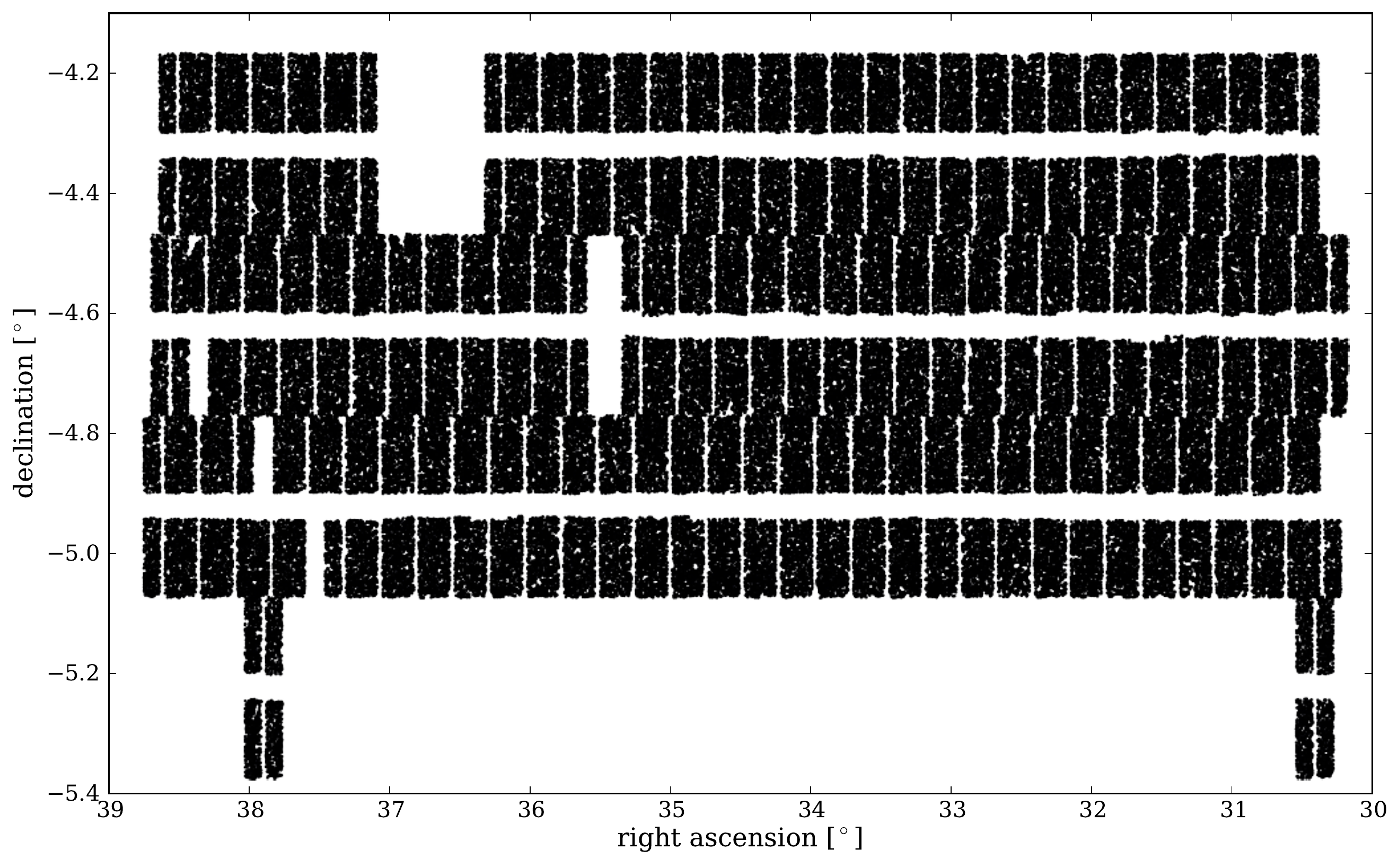}
\caption{The angular footprint of the VIMOS spectrograph across the W1 field of the VIPERS PDR-1 release \citep{Vipers}.  This angular selection represents an idealised VIPERS mask in which the spectroscopic success rate is unity and the target success rate is binary across the sky; see \protect \cite{sylxi} for further discussion.}
\label{angularfootprint}
\end{figure*}

This is a relation of significant practical importance as a Hankel transform may be evaluated in a single 1D FFT (\citejap{Hamilton}, FFTlog).  It is then possible to quickly transform between P$_{\ell}(k)$ and $\xi_{\ell}(\Delta)$ -- a fact we utilise to achieve both a $10^3-10^4 \times$ speedup and increased spectral resolution relative to a 3D FFT approach.  As a 3D FFT is $\sim 3 N^2 \times$ slower than a 1D FFT of mesh size $N$, for a given time, there is the potential for an improvement of $\sim 10^6$ in speed or a similar gain in resolution, alternatively a balance may be struck between the two.  As an additional benefit, when the problem is approached in configuration space there is no need to embed the survey in any effective volume and subsequently there is no resolution limit imposed by a corresponding fundamental mode.  A direct comparison of the resolution achieved is impractical as the optimised method we implement computes $P'_\ell(k)$ for logarithmically spaced intervals in $k$, as opposed to the linear spacing of the FFT.  This is a benefit in itself, due to the many decades in wavenumber over which $P'_{\ell}(k)$ is desired.  Consequently, the minimal amount of memory that is required allows aliasing associated with the 1D FFT to be confined to wavenumbers of no practical interest.  

\subsection{Practical details}
\label{practical}
Given that $\xi_{\ell}(\Delta)$ may be rapidly computed for an assumed RSD model, how is $P'_{\ell}(k)$ to be calculated?  In our approach, first the $Q_{q}(\Delta)$ are precomputed from a random catalogue bounded by the survey geometry, as described in \S\ref{sec:W2delta}.  At each point in the parameter space then the $\xi_{\ell}(\Delta)$ are computed by Hankel transformation (1D FFT) of $P_{\ell}(k)$ for $\ell=0,2,4, \cdots $.  For the monopole, the necessary linear combination is formed,
\[
\xi_0'(\Delta) = \xi_0 Q_0 + \frac{1}{5} \xi_2 Q_2 + \frac{1}{9} \xi_4 Q_4 + \frac{1}{13} \xi_6 Q_6 + \cdots, 
\label{monoxi}
\]
and an inverse Hankel transform (1D FFT) computes $P'_0$.

This relation is obtained by calculating the $A_{\ell, \ell'}^{q}$ coefficients that account for the weighted volume average given by eqn. (\ref{noangleavg}).  To illustrate this point, consider a simple case in which both density field and mask are composed of solely quadrupole terms.  Following eqns. (\ref{overautocorr}) \& (\ref{Bailey}) the $\mu$ dependence of the product is a linear combination of monopole, quadrupole and hexadecapole terms: $(1/5) L_0 + (2/7) L_2 + (18/35) L_4$, such that $\xi'_{2} = (2/7) \ \xi_2 Q_{2}$; in this case the masked monopole and quadrupole may be seen simply by inspection.  More generally a contribution to $P'_\ell(k)$ is generated by all $(q, \ell')$ terms of $Q_q$ and $\xi_{\ell'}$, due to the angular dependence of $L_{\ell} L_{\ell'}$ -- not least from leakage of the monopole power to the quadrupole via $Q_{2}$.  By calculating the coefficients $A_{2, \ell'}^{q}$ an explicit expression for the lowest order terms in $\xi_2'(\Delta)$ may be given:
\begin{eqnarray}
\xi_2'(\Delta) = \xi_0 Q_2 &+ \ \xi_2&\left( Q_0 + \frac{2}{7} Q_2 + \frac{2}{7} Q_{4} \right)  \nonumber \\
 		  &+ \ \xi_4&\left( \frac{2}{7} Q_2 + \frac{100}{693} Q_4 + \frac{25}{143} Q_{6} \right)   \nonumber \\ 
		  &+ \ \xi_6&\left( \frac{25}{143} Q_4 +  \frac{14}{143} Q_6 + \frac{28}{221} Q_8 \right) \nonumber \\
		  &+ \cdots &
\label{quadxi}
\end{eqnarray}
Equations (\ref{monoxi}) \& (\ref{quadxi}) together with the FFTlog implementation of eqn. (\ref{Hankelpair}) and its inverse suffice to calculate $P'_0(k)$ \& $P'_2(k)$.  As $\xi_{\ell}$ is non-zero for even $\ell$ only there is no dependence on $Q_q$ for odd-numbered $q$.  One might expect this expression to include $Q_q$ terms of arbitrary  order, but this is not the case as $L_{2} L_{\ell'}$ has only three non-zero Legendre coefficients for given $\ell'$.  We return to a test of the convergence rates of these expressions with respect to $\xi_{\ell}$ in Appendix \ref{convergence}. 

The required coefficients may be calculated to increasingly higher order to obtain a given precision, however the measurement of $Q_{q}$ becomes progressively noisier for $q \gg 1$; we comment on an approach to this problem in \S\ref{sec:W2delta}.  On large scales, $\Delta \gg 1 \mpcoh$, the series will be truncated by the Kaiser limit, $\xi_{\ell}(\Delta) = 0$ for $\ell > 4$ \citep{Hamilton92}, while on small scales the series will commonly be restricted by $Q_q(\Delta)$, which are typically much smaller than $Q_0$ for $\Delta < 10 \mpcoh$, see  Fig. \ref{WindowMultipoles} for instance. It is clear that of order $10$ one-dimensional FFTs are required for the prediction of $P_0'(k)$ \& $P_2'(k)$ if both $\xi(\Delta)$ and $Q(\Delta)$ are well approximated by terms for which $\ell$ \& $q \leq 6$.

Although the logical approach is to measure the multipole moments of the mask using a large random catalogue in the usual manner (see \S \ref{sec:W2delta}), they may instead be obtained by FFT as in other common approaches.  In this case, the configuration-space moments $Q_{q}(\Delta)$ required in our approach may be estimated by
\begin{eqnarray}
Q_{q}(\Delta)&=&i^{q} (2q +1) \int \frac{d^{3}k}{(2 \pi)^3} |\tilde{W}(\mathbf{k})|^{2} \ j_{q}(k \, \Delta) \ L_{q}(\mathbf{\hat{k}} \cdot \hat{\boldsymbol \eta}) \nonumber, \\
&\approx& i^{q} (2q +1) \sum |\tilde{W}(\mathbf{k}) |^2  \ j_{q}(k \, \Delta) \ L_{q}(\mathbf{\hat{k}} \cdot \hat{\boldsymbol \eta}).
\end{eqnarray}
Where the sum is restricted to the modes, $\mathbf{k}$, available with a 3D FFT.  To obtain this result we have used eqn. (\ref{FTwindow}), replaced plane with spherical waves and used the identity given by eqn. (\ref{eqn:Orthogonality}).  We performed a test of this final approximation and confirmed the validity of our approach with the following illustrative example.   

\subsection{A simple validity test}
\label{validity}
To test the validity of our approach we generated a set of 5000 realisations of $(500 \mpcoh)^3$ anisotropic Gaussian random fields, computed on a mesh with a cell size of $4 \mpcoh$, with a $(1 + \mu^2/2)P_R(k)$ power spectrum, to which a common 3D `mask' was applied and the resultant $P'_{\ell}(k)$ measured.  This anisotropy retains the lowest order terms of a Taylor expansion in $\mu$, while satisfying the requirements that the density field be both anisotropic and respect the symmetries of RSD -- additional terms being unnecessary for demonstrating that the effect of the mask may be correctly accounted for.  For the mask an independent Gaussian field provided the weighting, $W(\mathbf{x})$, necessary for evaluating eqn. (\ref{delta_prime}).  This weighting is purely for illustration; it is not representative of a realistic survey, the mask of which would commonly be separable into angular and radial dependencies.  The power spectrum of this mask is chosen for convenience to have the same $P(\mathbf{k})$ as the density field; this allows for the calculation of $Q_{\ell}(\Delta)$ by Hankel transformation of the known $P_{\ell}(k)$.  To increase the convergence rate of the mean ensemble power to the expectation the amplitude of each mode was assigned to the expectation value, as opposed to being drawn from an exponential distribution -- see \S 16.3 of \cite{CP}.  A comparison between that `observed' and the prediction of our method is shown in Fig. \ref{ConvldMultipoles}.  The results of this simple test confirm the validity of our approach, with excellent agreement obtained between the predictions of our method and the numerical simulations. 

\subsection{Obtaining $\boldsymbol Q_{\boldsymbol q}( \boldsymbol \Delta)$ with a pair counting approach}
\label{sec:W2delta}
In this section we outline a method for obtaining $Q_{q}(\Delta)$ by pair counting a random catalogue of constant number density that is bounded by the survey geometry and comment on the artefacts introduced by accounting for the mask with FFT based approaches. 

In Appendix \ref{paircounting} we provide a straightforward derivation that shows that if the randoms are distributed with constant number density $\overline{n}_s$ and are weighted by $W(\mathbf{x})$ then 
\[
\overline{RR}^{\mathrm{tot}}_{q}(\Delta) = \pi \overline{n}_s^2 \Delta^3 d(\ln\Delta) Q_{q}(\Delta).
\]
Here $\overline{RR}^{\mathrm{tot}}_{q}$ corresponds to distinct pairs binned by separation in logarithmic intervals, with each pair weighted by 
\[
\frac{(2q +1)}{2} L_{q}(\boldsymbol{\hat \Delta} \cdot \boldsymbol{\hat \eta}) W(\mathbf{x}_1) W(\mathbf{x}_2).  
\]
It is clear that $(\overline{RR}^{\mathrm{tot}}_{q}/ \Delta^3)$ differs only in amplitude from $Q_{q}(\Delta)$; the measured counts should then be similarly rescaled such that $(\overline{RR}^{\mathrm{tot}}_{0}/ \Delta^3) \mapsto 1$ for $\Delta \ll 1 \mpcoh$, as the renormalisation given by eqn. (\ref{ampcorr}) is equivalent to enforcing $Q_{0}(0)=1$.

This configuration-space pair counting approach has a number of advantages compared to an estimate of $|\tilde{W}(\mathbf{k})|^2$ by FFT:
\begin{figure}
\centering
\includegraphics[width=0.5\textwidth]{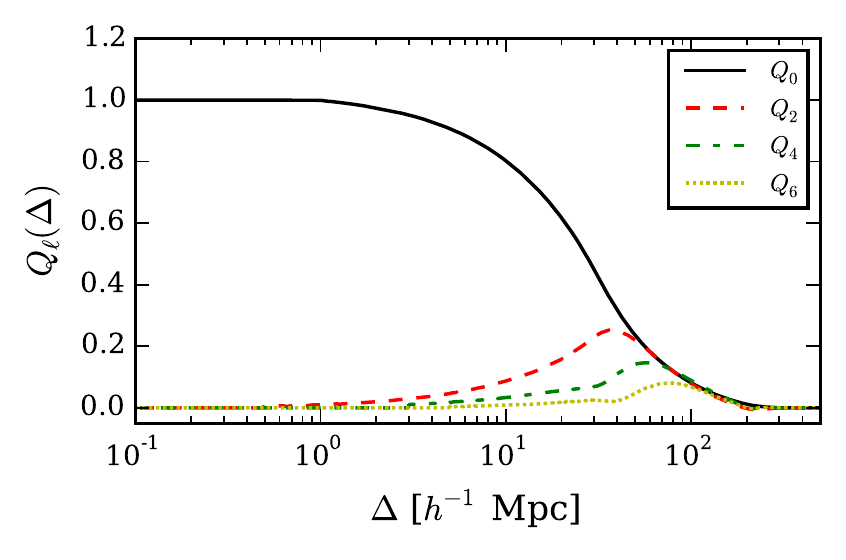}
\caption{Multipole moments of the VIPERS PDR-1 mask autocorrelation function, corresponding to the idealised angular selection shown in Fig. 2, for a redshift slice of $0.7<z<0.8$.  From this figure it is clear that the anisotropy of this VIPERS mask is significant for scales above $\simeq 10 \mpcoh$.}
\label{WindowMultipoles}
\end{figure}
\begin{enumerate}
\item  Large volume surveys are challenging with a memory-limited FFT due to the large volume required to embed the survey, which enforces a small fundamental mode and hence a small Nyquist frequency.  If the survey has small scale angular features then $|\tilde{W}(\mathbf{k})|^2$ at large $k$ may be large and the aliasing introduced by coarse binning may be significant.  
\item The integral constraint correction presented in \S \ref{sec:intcor} requires a robust estimation of $|\tilde{W}_{\ell}(k)|^2$ for $k \ll 1 \hompc$. 
A FFT is imprecise for this estimate due to the limited number of modes available in this regime.  In contrast, a Hankel transform of the pair counts yields a much higher resolution estimate and hence a more robust correction may be made; this is shown quite clearly in Fig. \ref{windowkmultipoles}. For further discussion see \S \ref{sec:intcor}.
\item The required pair counting is performed prior to the likelihood analysis and is easily optimised with the use of a k-d tree or a similar technique.  For $\Delta < 10 \mpcoh$, the relatively small number of pairs gives a noisy estimate but this regime can be rapidly remeasured with a higher density, by decreasing the maximum separation of nodes that are to be included.
\end{enumerate}

It is often the case that an additional weighting is applied to the surveyed volume, rather than simply a binary geometric factor.  For the commonly used FKP estimator (\citejap{FKP}, FKP), a further $\bar n(\mathbf{x})/(1 + \bar n(\mathbf{x}) P(\mathbf{k}))$ weighting is required; which may be seen by contrasting eqn. (2.1.6) of FKP with eqn. (\ref{convo}).  In this case the best approach is to generate a random catalogue with the same radial distribution function as the survey; each pair should then be weighted by
\[
\frac{(2q + 1)}{2} L_{q}(\boldsymbol{\hat \Delta} \cdot \boldsymbol{\hat \eta})/ \left [(1+ \bar n_1 P)(1 + \bar n_2 P) \right ].
\]
\begin{figure}
\centering
\includegraphics[width=0.5\textwidth]{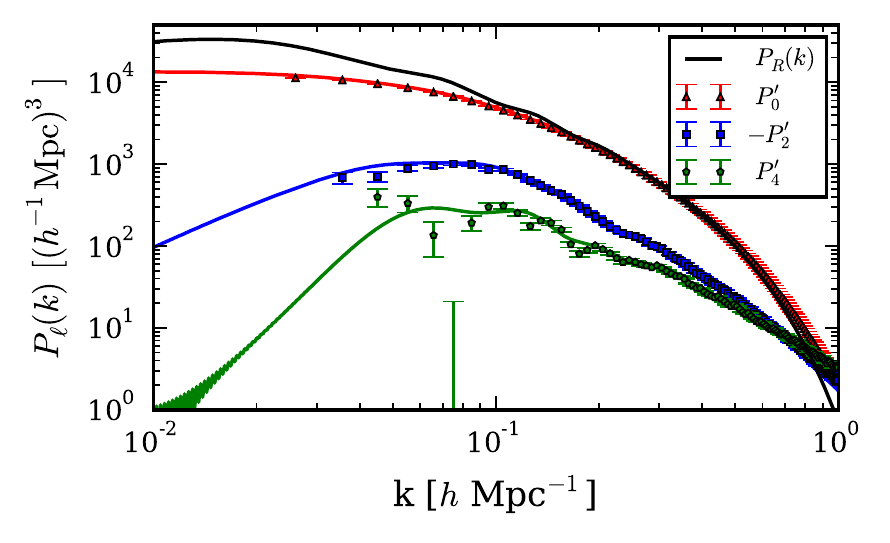}
\caption{Observed multipole moments for a set of Gaussian realisations with an \textit{isotropic} $P(k)$, to which we have applied a common mask corresponding to the W1 survey geometry for $0.7<z<0.8$. Here the observed quadrupole and hexadecapole moments result solely from the anisotropy of the survey mask, as given explicitly for the quadrupole by eqn. (\ref{quadxi}).  As previously, the convergence of the convolved to the unconvolved multipoles occurs for $k > 1 \hompc$.  While the slight oscillatory artefact visible in the hexadecapole for $k<0.02 \hompc$ is a result of the resolution choice for the 1D FFT of the FFTlog algorithm. Over the comparable scales, our predictions of the convolved moments (solid) are in excellent agreement with that observed in the realisations.}
\label{VIPERS_cnvld_hexpk}
\end{figure}
\begin{figure}
\centering
\includegraphics[width=0.5\textwidth]{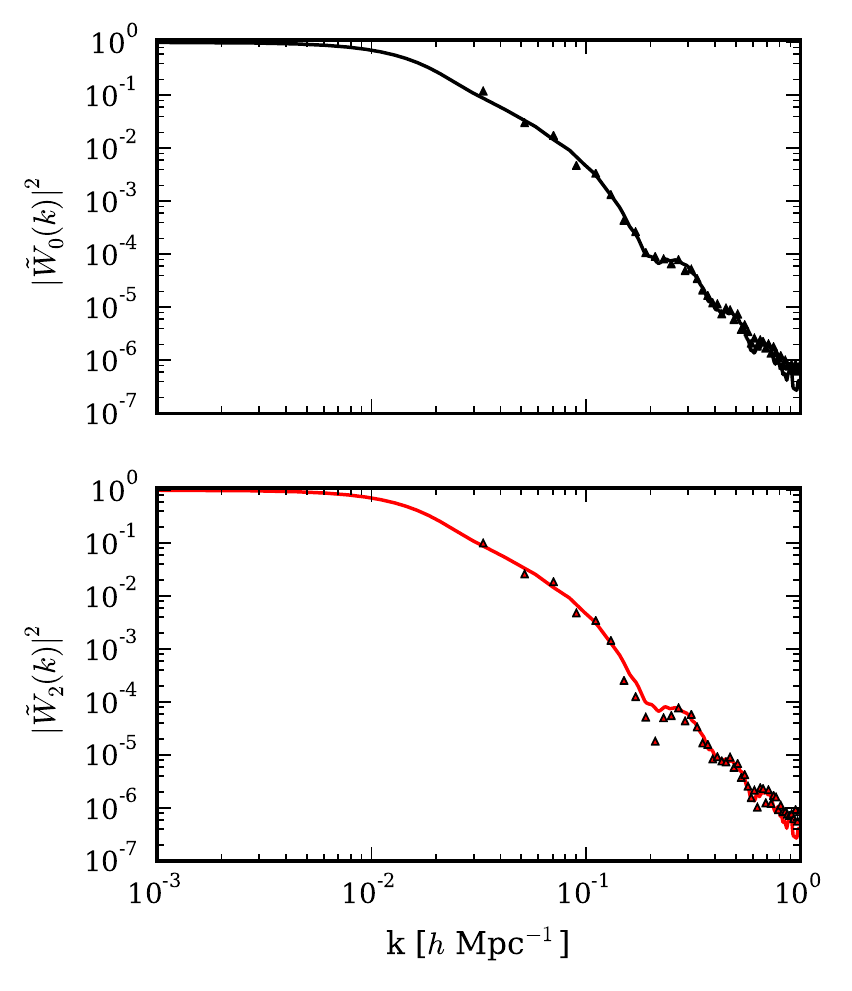}
\caption{A comparison of $|\tilde W_{0}(k) |^2$ \& $|\tilde W_{2}(k) |^2$  for the VIPERS W1 mask obtained via Hankel transform of the measured pair counts (solid) and by 3D FFT (triangles).  The FFT measurement is coarsely binned to suppress the statistical noise; in contrast, a Hankel Transform approach is independent of the fundamental period of any embedding volume. Note that where the integral constraint correction is largest, $k \ll 1 \hompc$, there are very few FFT modes to make a robust estimate of $|\tilde W_{\ell}(k) |^2$.}
\label{windowkmultipoles}
\end{figure}
\section{VIPERS: an application to a realistic survey geometry}
\begin{figure}
\centering
\includegraphics[width=0.5\textwidth]{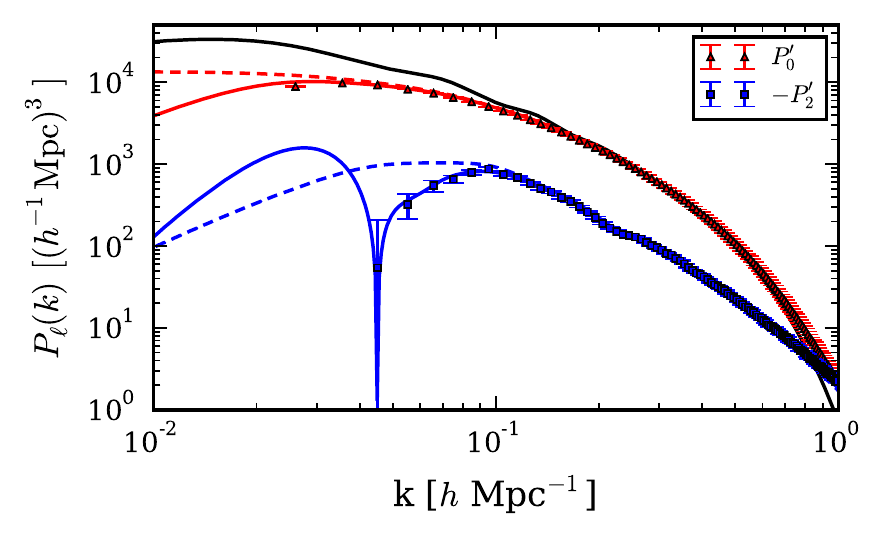}
\caption{Here we repeat Fig. \ref{VIPERS_cnvld_hexpk} but further assume that the surveyed volume is a fair sample of the density field, thereby introducing the integral constraint correction discussed in \S \ref{sec:intcor}.  We plot the monopole and quadrupole predictions before correction (dashed lines) and after (solid lines).  Clearly this is a significant impact on the observed quadrupole.  However, in a more realistic VIPERS analysis this could be reduced by estimating $\bar n$ from the combined W1 \& W4 CFHTLS-Wide fields, as opposed to this single field example.}
\label{VIPERShex_intcor}
\end{figure}
As a proof of principle we present a concrete application to a realistic test case, that of the W1 field of the VIMOS Public Extragalactic Redshift Survey (VIPERS) PDR-1 release \citep{Vipers}.  VIPERS is a large spectroscopic survey that has measured approximately 100,000 galaxies in the redshift range $0.5 < z < 1.2$.  Despite a significant volume, $\simeq 5 \times$ 10$^7$ ($h^{-1} \rm{Mpc}$)$^{3}$, and a relatively high sampling rate ($\simeq 40 \%$) VIPERS is afflicted by a complicated angular selection as shown in Fig. \ref{angularfootprint}.  The survey footprint comprises rows of pointings, each of which is made up of four quadrants separated by a central cross in which no spectra may be obtained.  This is in order to obtain the maximum volume possible and hence only a single pass is performed.  

This disjointed network presents a challenging test case for modelling the effect of the survey mask.  However, it should be noted that the convolution is dependent only on the mask autocorrelation function -- effectively the mask is smoothed by itself and hence the masked power spectrum is relatively insensitive to sharp angular features.  In Fig. \ref{WindowMultipoles} we show the lowest order moments, $Q_q(\Delta)$, of the VIPERS mask, when a volume limited sample spanning: $0.7<z<0.8$ is assumed; as a result the radial and angular dependence has comparable significance in determining the survey mask anisotropy.  From this figure it is clear that the anisotropy of this VIPERS mask is significant for scales above $\simeq 10 \mpcoh$.

Fig. \ref{VIPERS_cnvld_hexpk} shows the accurate prediction of the observed power, $P'_{0}$(k) \& $P'_{2}$(k) for a more realistic test case comprised of the application of this VIPERS mask to an isotropic Gaussian field, which illustrates the generation of the higher order multipoles due simply to the anisotropy of the survey mask. 

\section{Integral constraint correction}
\label{sec:intcor}
An additional correction is required for a masked density field: the background number density $\bar{n}$ is estimated from the finite survey volume and may differ from the true value due to clustering on wavelengths approaching the survey size.  Assuming the survey volume is a fair sample enforces the constraint: $\tilde \delta(\mathbf{0}) = 0$; as this false mean overdensity field is subject to the convolution detailed above a convolved spike centred on $\mathbf{k} = \mathbf{0}$ also contributes to the observed power.  To ensure $P^{\text{obs}}(\mathbf{0})=0$ the observed power is given by \citep{PeacockNicholson}
\[
P^{\text{obs}}(\mathbf{k}) = P'(\mathbf{k}) -  P'(\mathbf{0})  \ |\tilde{W} (\mathbf{k}) |^2. 
\]
Here $|\tilde{W}(\mathbf{k}) |^2$ is rescaled such that $|\tilde{W}(\mathbf{k}) |^2 \mapsto 1$ for $k \ll 1 \hompc$.  When the distant observer approximation is valid clearly this equates to
\[
P^{\text{obs}}_{\ell}(k) = P'_{\ell}(k) -  P'_{0}(0)  \ |\tilde{W} _{\ell}(k) |^2;
\label{intcor_eqn}
\]
in this case the rescaling is such that $|\tilde{W}_{0}(0) |^2 = 1$.  Given a measurement of  $Q_{\ell}(\Delta)$, $|\tilde{W}_{\ell}(k) |^2$ may be obtained by Hankel transformation; this approach allows for a much higher resolution measurement than with a 3D FFT -- free from the fundamental mode of any embedding volume.  In Fig. \ref{windowkmultipoles} we plot a comparison of the $|\tilde{W}_{\ell}(k) |^2$ obtained via these different approaches.  Note that the FFT estimate is coarsely binned in order to suppress the statistical noise present.

In Fig. \ref{VIPERShex_intcor} we repeat the example of the VIPERS mask applied to an isotropic density field but further assume that the mean density over the survey volume is a fair sample, thereby introducing an integral constraint to be corrected for.  The impact on the quadrupole is significant for this simplified case, but in a more realistic analysis this impact could be reduced by estimating $\bar n(z)$ from the combined W1 \& W4 CFHTLS-Wide fields observed by VIPERS.

\section{Complete impact of the VIPERS mask}
\begin{figure*}
\centering
\includegraphics[width=0.9\textwidth]{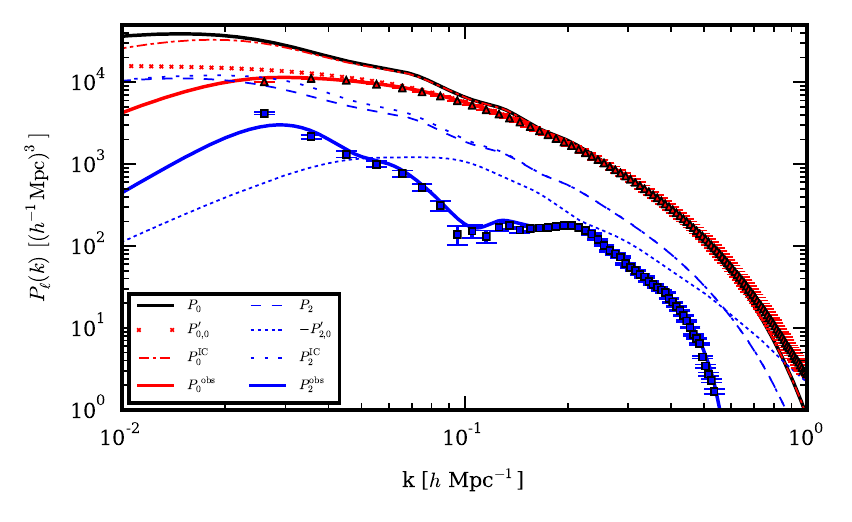}
\caption{This figure illustrates a breakdown of all necessary corrections due to the VIPERS PDR-1 W1 mask for an anisotropic density field with a $(1 + \mu^2/2)P_R(k)$ power spectrum, such that $P_0(k) = (7/6) P_R$ and $P_2(k) = (1/3) P_R$ in the absence of the mask.  Firstly, the isotropic component of the density field yields a distorted monopole, ${P'}_{0,0}$, in the presence of the mask and contributes significantly to the quadrupole, ${P'}_{2,0}$.  This leaked component may dampen the intrinsic quadrupole, depending on the relative anisotropy of the density field and mask.  We apply the appropriate integral constraint correction to the unmasked power for illustration, finding $P_{0}^{\rm{IC}}$ \& $P_{2}^{\rm{IC}}$.  The combination of these corrections gives the complete impact of the VIPERS mask on the monopole and quadrupole (solid), which are in excellent agreement with the mean of the Gaussian realisations.}
\label{intcor}
\end{figure*}

All necessary corrections for the VIPERS PDR-1 W1 mask are summarised in Fig. \ref{intcor}.  Here we analyse an intrinsically anisotropic density field with a $(1 + \mu^2/2)P_R(k)$ power spectrum, such that $P_0(k) = (7/6) P_R$ (solid black) and $P_2(k) = (1/3) P_R$ (blue long-dashed) in the absence of the survey mask -- this quadrupole-to-monopole ratio is equivalent to that obtained from the Kaiser model for $\beta \simeq (1/4)$.  In addition, a decomposition of the individual contributions are shown.  Firstly, the intrinsic \textit{isotropic} component of the density field, when masked, yields a distorted monopole: ${P'}_{0,0}$, and contributes significantly to the quadrupole, ${P'}_{2,0}$;  this leaked component may dampen the intrinsic quadrupole, depending on the relative anisotropy of the density field and mask.  We apply the appropriate integral constraint correction, the second term on the right of eqn. (\ref{intcor_eqn}), to the unmasked power for illustration; this is shown as $P_{0}^{\rm{IC}}$ \& $P_{2}^{\rm{IC}}$ respectively.  The combination of these corrections gives a prediction of the complete impact of the VIPERS mask on the observed multipoles (solid), which is in excellent agreement with that measured from the ensemble of Gaussian realisations. 
\section{Validity for finite-angle surveys}
\label{wide-angle}
Our approach assumes the validity of the distant observer approximation for the modelling, in which the variation of the line-of-sight across the survey is assumed to be negligible.  This is as opposed to the measurement on the data, which may (approximately) account for finite-angle effects (\citejap{Yamamoto}, \citejap{Bianchi}, \citejap{ScoccimarroWA}, \citejap{Slepian_WA}).  In the distant observer approximation the redshift-space $P(\mathbf{k})$ possesses the symmetries outlined in the introduction; if this assumption is relaxed and pairs with a finite angular separation are included in the analysis then the redshift-space $\xi$ is dependent on the triangular configuration formed by a given pair and the observer \citep{SphericalRSD}.  As there remains a statistical isotropy about the observer, any such configuration may be rotated into a common plane \citep{Szalay97}, following which the remaining degrees of freedom are vested solely in the triangular shape.  Alternative parametrisations of this configuration are possible, with a subsequent ambiguity in the definition of the `line-of-sight'.  One possibility is to define $\boldsymbol \eta$ as that bisecting the opening angle; in this case the triangle is fully defined by $\mu$, the pair separation and the opening angle: $\theta$ (see Fig. 1 of \citejap{Yoo} and discussion therein).  

It is clear that the finite-angle redshift-space $\xi$ does not possess the symmetries we exploit in the distant-observer limit, $\theta \mapsto 0 \deg$.  As such our method must be applied with some care to surveys with a large sky coverage.  While this is currently a limitation, the median redshift of future surveys will be considerably larger and the modal opening angle of relevant pairs will be $\simeq 4-6 \deg$ as compared to the $\simeq 20 \deg$ of current surveys, e.g. the Sloan Digital Sky Survey \citep{Yoo}.  In fact, Fig. 7 of \cite{Yoo} shows that the systematic error introduced by assuming the distant observer approximation in the modelling is negligible for both Euclid and DESI, provided the redshift evolution of both the density field and galaxy bias is correctly accounted for.  Moreover, the applicability of our approach to current finite-angle surveys is evidenced by recent applications to the BOSS DR12 dataset (\citejap{Beutler_2016A}, \citejap{Beutler_2016B}, \citejap{Zhao}).  In any case, a practical perspective is to accept that any bias introduced by using an approximate model, such as one which assumes the distant observer approximation, may be calibrated with numerical simulations and a resulting correction applied to the final data analysis.  Finite-angle effects are merely one instance where this approach may be taken.   

\section{Conclusions}
The effect of the survey mask represents the largest systematic difference between the observed large-scale power spectrum and that predicted by fundamental physics.  This work presents a new forward modelling approach for predicting the redshift-space power spectrum multipole moments in light of this effect.  By exploiting the symmetries of the redshift-space correlation function in the distant observer approximation, we derive a masked-field generalisation of the known Hankel transform relation between the multipole moments in real and Fourier space.  As a Hankel Transform may be computed with a 1D FFT, this implementation is $10^3 - 10^4 \times$ faster than other common approaches and achieves a higher spectral resolution, while keeping the introduction of FFT based artefacts to a minimum.  These benefits are especially relevant for large volume surveys with sharp angular features.  We also note that a similar solution has been suggested for calculating the convolutions necessary for the Eulerian 1-loop power spectrum \citep{Schmittfull}; the combination of this with our approach would allow for a particularly rapid prediction for the the large-scale masked power spectrum. 

We then describe and validate an approach for obtaining the required multipole moments of the mask autocorrelation function, $Q_q(\Delta)$, by pair counting a random catalogue bounded by the survey geometry.  By accounting for the mask in configuration space, rather than Fourier space, a more robust integral constraint correction may be made.  This approach allows for greater physical insight into the impact of the mask anisotropy and, with a k-d tree or similar, the calculation is conveniently optimised.   Although other approaches have achieved similar speeds, e.g. the `mixing matrix' approach of \cite{Blake}, this formulation suggests a physically motivated compression of the mask into the multipole moments of the autocorrelation function, $Q_q(\Delta)$.  With these in hand, any $P_{\ell}'(k)$ may be calculated with no restrictions on the wavenumber range or resolution.  This is in contrast to the mixing matrix, which achieves a compression by restricting the allowed modelling to a limited number of $P_{\ell}'(k)$, for a predefined set of wavenumber bins. 

A concrete application to the survey geometry of the VIPERS PDR-1 W1 field is presented.  We show that the power spectrum multipole moments can be accurately predicted and thereby provide a proof of principle for our approach.  Amongst other corrections, a significant quadrupole is generated by the quadrupole component of the autocorrelation function of the survey mask, as has been noted for other surveys \citep{Beutler}.  Although our method is limited to the distant observer approximation, and must be applied with care to large solid angle surveys, this issue will be mitigated by the larger median redshift of future surveys.  In practice, any small error introduced by finite-angle effects may be calibrated with numerical simulations and an appropriate correction applied to the final data analysis. 

The machinery constructed in this work should prove valuable in the RSD analyses of future galaxy redshift surveys such as VIPERS, eBOSS, DESI \& Euclid, allowing the systematic error introduced by the mask anisotropy to be rapidly corrected for.  This serves as an important step towards providing robust constraints on modified gravity theories based on the observed linear growth rate of density fluctuations. 
\section*{Acknowledgements}
MJW would like to acknowledge the recent passing of his mother, Dorothy Wilson, without whom this work would not have been possible.  We would also like to thank Florian Beutler, Gigi Guzzo, Chris Blake and Ben Granett for providing helpful comments. 
MJW was supported by a STFC PhD studentship.  JAP was supported by ERC grant number: 670193.  MJW and SdlT acknowledge the support of the OCEVU Labex (ANR-11-LABX-0060) and the A*MIDEX project (ANR-11-IDEX-0001-02) funded by the `Investissements d'Avenir' French government program managed by the ANR.
\appendix
\section{The Kaiser-Lorentzian model}
\label{sec:appA}
A brief summary of the power spectrum multipole moments for the Kaiser-Lorentzian dispersion model is given below.  The variables used are consistent with those defined in the discussion following eqn. (\ref{DispersionModel}).  The monopole and quadrupole moments are:
\begin{align}
\frac{P_0(k)}{P_R(k)} &=M_0(\kappa) + 2 \beta M_2(\kappa) + \beta^2 M_4(\kappa), \nonumber \\
\frac{2}{5} \frac{P_2(k)}{P_R(k)} &= -M_0 + (3 - 2 \beta) M_2  + \beta (6 -\beta) M_4 + 3 \beta^2 M_6, \nonumber \\
\end{align}
where
\begin{equation}
M_n = \int_0^1 \frac{\mu^n d \mu}{1 + \frac{1}{2} k^2 \sigma_p^2 \mu^2}, \\
\end{equation}
and, for $\kappa = k \sigma_p$, 
\begin{align}
&M_0(\kappa) = \frac{\sqrt{2}}{\kappa} \arctan(\kappa /  \sqrt{2}),&  \nonumber \\
&M_2(\kappa) = \frac{2}{\kappa^3} \left( \kappa - \sqrt{2} \arctan(\kappa /  \sqrt{2}) \right),& \nonumber \\
&M_4(\kappa) = \frac{2}{\kappa^5} \left( -2 \kappa + \frac{\kappa^3}{3} + 2\sqrt{2} \arctan(\kappa /  \sqrt{2}) \right),& \nonumber \\
&M_6(\kappa) = \frac{2}{\kappa^7} \left( 4 \kappa - \frac{2}{3} \kappa^3 + \frac{\kappa^5}{5} -4 \sqrt{2} \arctan(\kappa /  \sqrt{2}) \right).& 
\raisetag{-.25em}
\end{align}
\section{Derivation of eqn. (12)}
Here we reproduce the identities from Appendix (B3) of \cite{Beutler}, which are required for the derivation of the relation:
\[
\frac{(2\ell + 1)}{2} \int d(\hat{\mathbf{k}} \cdot \hat{\boldsymbol \eta}) \int \frac{d \phi_k}{(2 \pi)} L_{\ell}(\hat{\mathbf{k}} \cdot \hat{\boldsymbol \eta}) L_{\ell'} (\hat{\mathbf{k}} \cdot \hat{\boldsymbol \Delta})  = \delta^{K}_{\ell \ell'} L_{\ell}( \hat{\boldsymbol \Delta} \cdot \hat{\boldsymbol \eta}). \label{B1}
\]
Starting with the definition of the Legendre polynomials in terms of spherical harmonics,
\[
L_{\ell}(\hat{\mathbf{k}} \cdot \hat{\boldsymbol \eta}) = \frac{4 \pi}{(2 \ell +1)} \sum_{m = -\ell}^{\ell} Y_{\ell m}(\hat{\mathbf{k}}) Y^{*}_{\ell m}(\hat{\boldsymbol \eta}),
\]
together with the orthogonality relation, 
\[
\int d(\hat{\mathbf{k}} \cdot \hat{\boldsymbol \eta}) \int d \phi_k \, Y_{\ell m}(\hat{\mathbf{k}}) Y^{*}_{\ell' m'} (\hat{\mathbf{k}}) = \delta^{K}_{\ell \ell'} \delta^{K}_{m m'},
\]
eqn. (\ref{B1}) results from a straightforward derivation.
\section{Convergence of $\boldsymbol P'_{\boldsymbol 0}$ \& $\boldsymbol P'_{\boldsymbol 2}$}
\label{convergence}
\begin{figure}
\centering
\includegraphics[width=0.5\textwidth]{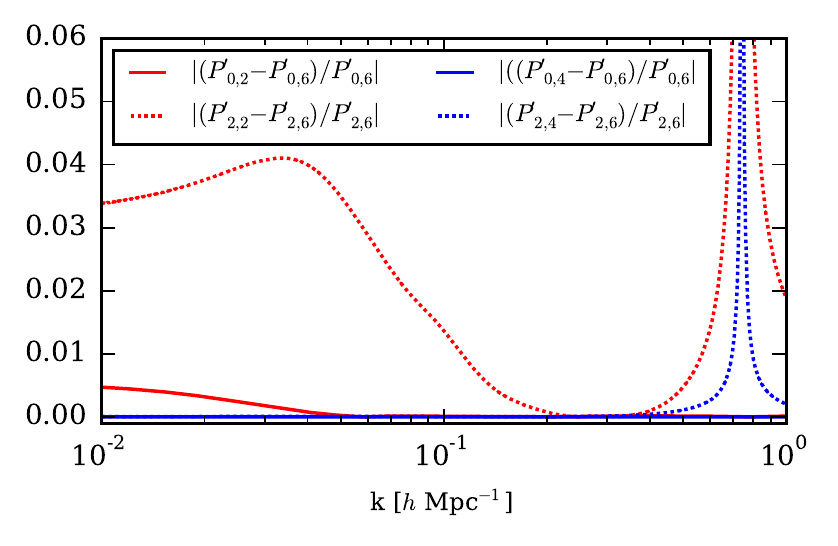}
\caption{This figure illustrates the convergence rate of $P'_{\ell}(k)$ with respect to the expansion in $\xi_{\ell'}$ given in eqns. (\ref{monoxi}) \& (\ref{quadxi}).  Here $P'_{0,p}$ denotes the predicted monopole power spectrum for the masked density field when the expansion is truncated at $\xi_{p}$ (inclusive), and similarly for the quadrupole, $P'_{2, p}$.  We analyse the VIPERS PDR-1 geometry and assume $(\beta, \sigma_p) = (0.5, 5.0)$, corresponding to a conservative choice of $\sigma_p$.  For $\sigma_p \ll 1 \mpcoh$ the Kaiser model is recovered and the series formally converges with $\xi_4$.  From this figure it is clear that the inclusion of the hexadecapole terms is sufficient for obtaining subpercent precision on the masked multipoles.  There is a divergence in the fractional error of the quadrupole at $k = 0.75 \hompc$ as $P'_{2,6}$ passes through zero at this point.}
\label{fig_convergence}
\end{figure}

Fig. \ref{fig_convergence} shows the result of a test of the convergence rate of the expressions for $\xi_{\ell}'$ given in \S \ref{practical}.  In this figure $P'_{0,p}$ denotes the predicted monopole power spectrum for the masked density field when the expansion is truncated at $\xi_{p}$ (inclusive), and similarly $P'_{2,p}$ denotes the quadrupole.  We analyse the VIPERS PDR-1 mask and assume $(\beta, \sigma_p) = (0.5, 5.0 \mpcoh)$, corresponding to a conservative choice of $\sigma_p$ for which the series is expected to be converging relatively slowly.  For small values of $\sigma_p$ the Kaiser model is recovered and $\xi_{\ell'}(\Delta) =0$ for $\ell'>4$.  From this figure it is clear that the inclusion of the hexadecapole terms is sufficient for obtaining subpercent precision on the masked multipoles.  The fractional error of the quadrupole diverges at $k=0.8 \hompc$ as $P'_{2,6}$ passes through zero at this point.  To calculate the higher order multipole moments we make use of an implementation of \cite{DeMicheli}, which enables us to obtain the first $N$ multipole moments with a 1D FFT of size $N$, for one value of $k$.
\section{A pair counting approach}
\label{paircounting}
Consider the available distinct pairs of a random catalogue of constant number density, $\overline n_s$, that populates two volumes, $dV_1$ \& $dV_2$.  With the application of weights $W(\mathbf{x})$, the weighted pair count is given by 
\[
RR(\mathbf{x}_1, \mathbf{x}_2) = \frac{1}{2} \, \overline{n}_s^2 dV_1 dV_2 W(\mathbf{x_1}) W(\mathbf{x_2}). 
\]
If $dV_2$ is taken to be centred on $\boldsymbol x_2 = \boldsymbol x_1 + \boldsymbol \Delta$ and the total such pairs are counted over the surveyed volume, $\int dV_1$, then  
\[
RR^{\mathrm{tot}}(\boldsymbol \Delta) = \frac{1}{2} \overline n_s^2 Q(\boldsymbol \Delta) dV_2.
\]
By first applying a $(2q+1) L_q(\boldsymbol{\hat \Delta} \cdot \boldsymbol {\hat \eta})/2$ weighting to each pair and taking $dV_2$ as a narrow shell of width $d(\ln \Delta)$ centred on $\boldsymbol x_1$, the total number of weighted pairs summed over the surveyed volume is
\[
\overline{RR}^{\mathrm{tot}}_{q}(\Delta) = \frac{1}{2} \overline{n}_s^2 2 \pi \Delta^3 d(\ln \Delta) Q_q(\Delta).
\]
Significantly,  $\overline{RR}^{\mathrm{tot}}_{q}$ differs only in amplitude from $Q_q(\Delta)$; for each $q$, the counts should be similarly rescaled such that $\overline{RR}^{\mathrm{tot}}_{0}/\Delta^3 \mapsto 1$ for $\Delta \ll 1 \mpcoh$, as the renormalisation given by eqn. (\ref{ampcorr}) is equivalent to enforcing $Q_0(\boldsymbol 0) = 1$.

\setlength{\bibhang}{2.0em}
\setlength\labelwidth{0.0em}

\bibliography{biblio}
\end{document}